\documentclass[10pt,conference,letterpaper]{IEEEtran}
\IEEEoverridecommandlockouts
\ifCLASSINFOpdf
\else
\fi
\usepackage{graphicx}
\usepackage{amsmath}
\usepackage[ruled]{algorithm2e}
\usepackage{subfigure}
\usepackage{subfig}
\usepackage{booktabs}
\usepackage{tabu}

\usepackage{txfonts}
\usepackage{amssymb}
\usepackage{amsmath}
\usepackage{algpseudocode}
\usepackage{graphics}
\usepackage{epsfig}
\usepackage{cases}

\begin{document}
\title{Predictive Pre-allocation for Low-latency Uplink Access in Industrial Wireless Networks}
\author{\IEEEauthorblockN{Mingyan Li, Xinping Guan, Cunqing Hua, Cailian Chen and Ling Lyu}\\
\IEEEauthorblockA{Department of Electronics, Shanghai Jiao Tong University,\\
and Key Laboratory of System Control and Information Processing, Ministry of Education of China \\
Shanghai 200240, P.R. China \\}
Email:\{limy2015, xpguan, cqhua, cailianchen, sjtulvling\}@sjtu.edu.cn
\thanks{This work was supported in part by NSF of China under Grants 61521063, 61622307, U1405251, 61371085 and 61603251, the National High Technology Research and Development Program of China (863 Program) (no. 2015AA01A702).}}

\maketitle

\begin{abstract}
Driven by mission-critical applications in modern industrial systems, the 5th generation (5G) communication system is expected to provide ultra-reliable low-latency communications (URLLC) services to meet the quality of service (QoS) demands of industrial applications. However, these stringent requirements cannot be guaranteed by its conventional dynamic access scheme due to the complex signaling procedure. A promising solution to reduce the access delay is the pre-allocation scheme based on the semi-persistent scheduling (SPS) technique, which however may lead to low spectrum utilization if the allocated resource blocks (RBs) are not used. In this paper, we aim to address this issue by developing DPre, a predictive pre-allocation framework for uplink access scheduling of delay-sensitive applications in industrial process automation. The basic idea of DPre is to explore and exploit the correlation of data acquisition and access behavior between nodes through static and dynamic learning mechanisms in order to make judicious resource per-allocation decisions. We evaluate the effectiveness of DPre based on several monitoring applications in a steel rolling production process. Simulation results demonstrate that DPre achieves better performance in terms of the prediction accuracy, which can effectively increase the rewards of those reserved resources.
\end{abstract}

\IEEEpeerreviewmaketitle

 \vspace{-3pt}
\section{Introduction}
\vspace{-3pt}
    In order to promote the revolution of Internet of Things (IoT) connectivity, the coming 5G communication system is expected to expand traditional industrial informatics and automation systems into much broader contexts. As one of the most important scenarios in machine type communications (MTC), URLLC is driven by those mission-critical applications which require robust and timing predictable transmissions~\cite{5Glow}\cite{SPAL}. The reaction time in these applications is normally on the order of millisecond for real-time interaction, such as industrial automation~\cite{chencialian}, intelligent transportation systems and smart grid~\cite{maosiwen}. In particular, industrial automation typically consists of many automated manufacturing steps and involves many closed-loop industrial wireless sensor-actuator networks. Thus, the timely delivery of data is critical for process monitoring and control since missing a deadline may severely degrade the control quality, even worse, resulting in serious economic losses and safety problems.

    As a promising paradigm for industry automation, the 5G communication systems bring not only the benefits of wireless communication, such as deployment scalability, easier installation and maintenance, etc.~\cite{5Gad}, but also the inherent advantages for supporting deterministic medium access techniques. Unfortunately, traditional downlink-centric cellular networks fail to guarantee the strict timeliness in industrial automation due to the dynamic access scheme. In the conventional dynamic access procedure, a device should send a scheduling request (SR) to the base station (BS) to inform its intention for uplink data transmission. After receiving the scheduling grant (SG) from the BS, the device should send back a buffer state report (BSR), and finally the device is allowed to transmit according to the assigned resource blocks. As a result, this complex signaling procedure will lead to large access delay~\cite{factory}.

    As a solution, in LTE Release 13, an instant uplink access (IUA) scheme was proposed based on the SPS technique~\cite{R13}, whereby the uplink resources are assigned to the devices in advance without explicit SR-SG signalling procedure, which is suitable for mission-critical applications in industrial automation. Note that SPS is originally designed for VoIP services, which can be perfectly scheduled since their transmission rate is generally fixed and known a priori. However, the traffic of MTC devices with different QoS requests is typically sporadic and variable in industry. Therefore, access latency will be reduced at the expense of spectrum resources if SPS is applied for the uplink scheduling in MTC without enhancements~\cite{SPSv}. As the solution, some efforts have been made to deal with diverse QoS requirements through clustering~\cite{AGTI}\cite{AGTILQ}\cite{AGTI3}, and coarse granularity of pre-assigned resources via adaptive allocation by BSR, or reusing vacant dedicated resources through device-to-device (D2D) technique~\cite{ASPS}\cite{delegation}. However, some of these solutions may lead to extra access latency due to the additional overhead during information exchange. In addition, some of these works are based on the stochastic assumption of the arrival and service processes, which is not necessarily accurate for modeling the industrial traffic.

    Contrary to traditional mobile broadband devices that access the network independently, a salient feature of industrial automation is that one event-triggered MTC device may increase the probability that other devices in the vicinity also generate data in quick succession~\cite{predictive8}. This has been exploited to design a predictive resource allocation scheme~\cite{Predictive}, which proactively assigns the uplink resources to the neighboring devices according to the distance between them. However, simply pre-allocating resources to the neighbors is not necessarily optimal, since there may exist irrelevant nodes in the vicinity, and the scheduled nodes may not be triggered due to network dynamics such as thread burst/shutdown, node duty cycle, dwindling battery reserves, and node failure~\cite{dynamics}. In addition, it is likely to reserve for outdated data if the diverse QoS requirements are not taken into account, which is wasteful since the RBs available for reservation are limited.
    \begin{figure}[t]
            \centering
            \includegraphics[height=8.2cm, width=0.48\textwidth]{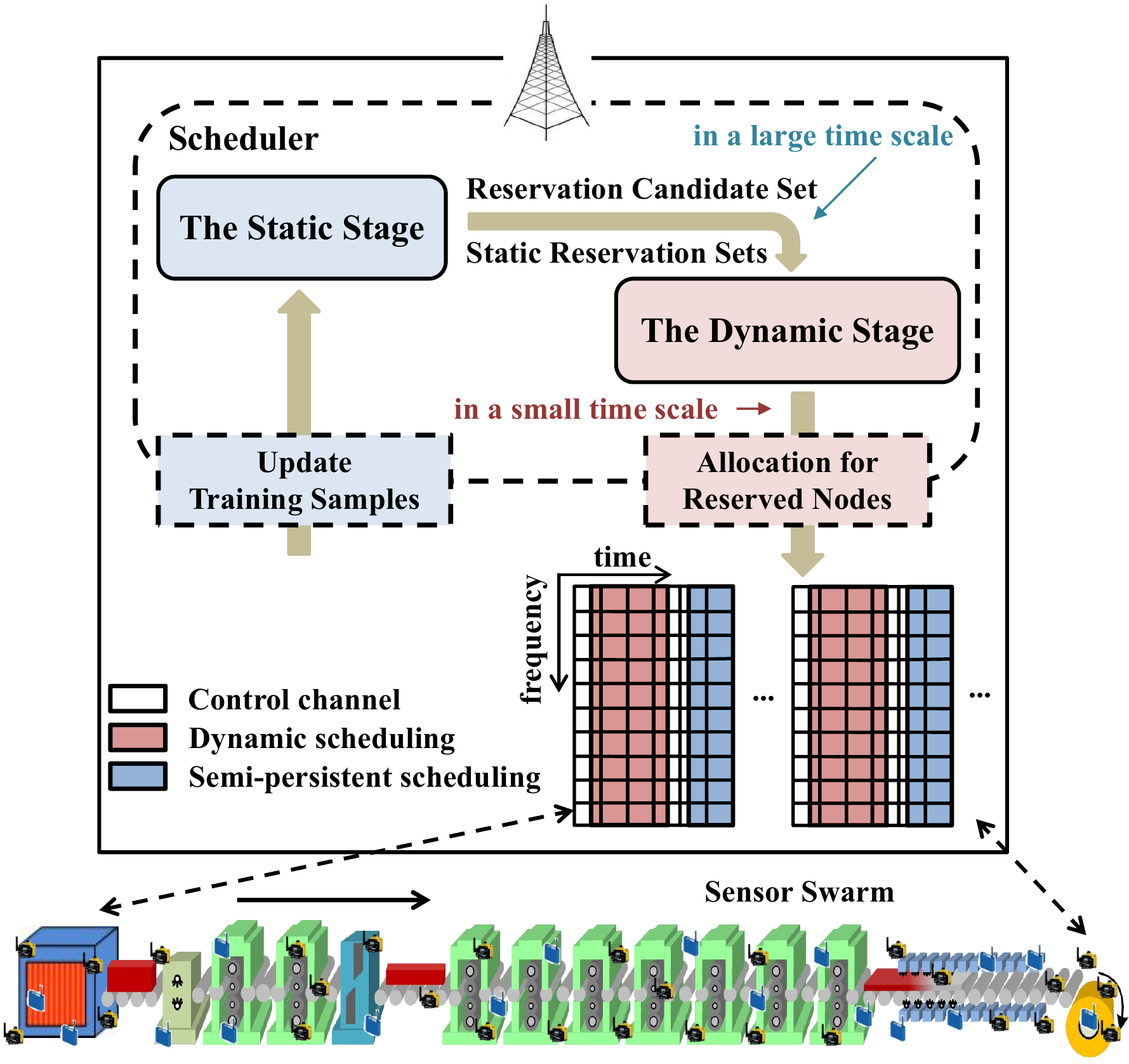}
            \vspace{-3pt}
            \caption{The System Architecture of DPre}
            \label{layer}
            \vspace{-22pt}
    \end{figure}

    Therefore, a more intelligent prediction strategy should be developed to boost the rewards of pre-allocated resources. This predictive pre-allocation problem with no prior knowledge of traffic arrival process is akin to the famous adversarial multi-armed bandit problem (AMAB)~\cite{monitoring}\cite{activelearning}, in which the BS sequentially learns the action profile and concurrently makes decisions to reserve for $M$ out of $N$ non-identical devices in a sequence of trials so as to maximize its payoff. However, with enormous available choices (arms) in our problem, where $M\ll N$ due to the massive devices deployed, it is indeed quite pertinent to eliminate unnecessary exploration.

    In this paper, we propose DPre, a predictive pre-allocation framework for low-latency uplink access scheduling in industrial process automation, where sensors are sequentially triggered according to monitored objects. DPre consists of a static and a dynamic stage. The static stage attempts to scale down the exploration space of all nodes to static reservation sets learned through the access history. Then network dynamics and diverse QoS requirements are considered in the dynamic stage based on the AMAB scheme, which explores nodes partially according to the correlation information obtained in the static stage and then makes pre-allocation decisions.


      The remainder of this paper is organized as follows. An overview of the DPre framework is presented in Section~\ref{overview}. The details of static and dynamic stages are discussed in Section~\ref{static} and~\ref{dynamic}, respectively. Then, we evaluate the performance of DPre in Section~\ref{performanceevaluation}. Finally, the related work is summarised in Section~\ref{relatedwork} and the conclusion is drawn in Section~\ref{conclusion}.
      \vspace{-3pt}
\section{Overview}~\label{overview}\vspace{-0pt}
The architecture of DPre for an industrial wireless network is illustrated in Fig~\ref{layer}. The network consists of a single BS and a set of $\mathcal{N}=\{1,..,N\}$ sensor nodes for process monitoring. All these nodes are directly served by the BS~\cite{5Glayer} with the single carrier frequency division multiple access (SC-FDMA) technique for the uplink. The BS is in charge of the assignment of resource blocks (RBs), where one RB denotes a series of time-frequency domains radio resources and is sufficient for most packet transmissions due to the small data feature in MTC~\cite{AGTI}.
In practice, the number of RBs available for pre-allocation in each transmission time interval (TTI), denoted by $Nres$, is much smaller than the number of nodes (i.e., $Nres\ll N)$ and the rest of RBs is for dynamic access. Therefore, the main objective of the BS is to reserve for the appropriate subset of sensor nodes from $\mathcal{N}$ in each TTI based on the trigger correlation of nodes, which can be achieved through our static and dynamic stages. Workflows of these two stages are outlined in Algorithm 1 and 2, respectively.

    \begin{algorithm}[b]
    \caption{Static Stage Workflow}
    \label{A1}
    \KwIn{ History access samples $Q_p = \{(\mathbf{x_i},y_i)\}_{i=1}^{I}$}
    \KwOut{Reservation candidate set $\Psi$; \\
           \ \ \ \ \ \ \ \ \ \ Corresponding static reservation sets $\mathcal{R}$;}
    \For{$ p = 1,2,3...$}
    {$\mathbf{\theta_p}$ = train$(Q_p)$ \;
     $\Psi$ = $\emptyset$ ; $\mathcal{R}=\emptyset $;\\
     \For{$i = 1,2,3...,I$}{
        Compute the correlation between $y_i$ and each of its feature nodes $x\in \mathbf{x_i}$, denoted as $\mathcal{E}(x,y_i)$ \;
        \If{$\Pi^{static}(\mathcal{E}(x,y_i))$}{
         $\Psi$ = $\Psi \cup \{y_i\}$ \;
        Select the static reservation set $\mathcal{R}(y_i)$ \;
        }
     }
     Perform the dynamic stage based on $\Psi$ and $\mathcal{R}$ \;
     Update($Q_{p+1}$) \;
     }
\end{algorithm}
Specifically, the aim of the static stage is to firstly decide the reservation candidate set $\Psi$, which consists of the nodes whose correlated nodes deserve pre-allocating transmissions. Then for each reservation candidate $y$ in $\Psi$, a static reservation set $\mathcal{R}(y)$ is selected from its potential correlated nodes. Through supervised learning, the BS calculates the correlation between nodes based on the history access samples, as shown in Algorithm~\ref{A1}. Then for each node $y$, a static strategy $\Pi^{static}$ is performed based on the correlation metric to determine whether $y$ is eligible to be a reservation candidate. If so, a static reservation set will be selected for it. Note that the BS keeps collecting access information during the dynamic stage in order to update $\Psi$ and $\mathcal{R}$ in the next step, which is performed in a large time scale (i.e., on the order of hours). The details will be introduced in Section~\ref{static}.
    \begin{algorithm}[b]
    \caption{Dynamic Stage Workflow}
    \label{A2}
    \KwIn{$\Psi$,$\mathcal{R}$;}
    \KwOut{The set of reserved nodes $\Omega$ for each TTI;}
    \For{$ t = 1,2,3...$}
    { $\Omega = \emptyset $ \;
      Collect the reservation candidates in this TTI:
      \begin{equation}
      \Theta_t = \{ y \mid y \in \mathcal{S}_{t-1} \cup \mathcal{C}_{t-1}, y \in \Psi \} \; \nonumber
       \end{equation}
      Decide the number of RBs $\delta_y$ assigned to each node $y \in \Theta_t$  \;
      \For{$y = 1,2,3...$}{
         Perform $\Pi^{dynamic}$ to choose the reserved nodes $\mathbf{k_y} \subset \mathcal{R}(y)$ in this TTI\;
         $\Omega = \Omega \cup \mathbf{k_y}$ \;
        }
      Pre-allocate RBs for the nodes in $\Omega$\;
      Find $\mathcal{S}_t$ and $\mathcal{C}_{t}$ in this TTI\;
      Renew the access samples Q \;
     }
\end{algorithm}

In the dynamic stage, the BS continuously updates the reserved node set $\Omega$ at the beginning of each time step $t$ and then broadcasts the reservation information to all nodes. Here, the time step $t$ corresponds to one TTI. we assume that each node in $\Omega$ can successfully receive the reservation information and transmit data using the allocated RB as long as it is triggered. Specifically, in each TTI $t$, the BS firstly collects the reservation candidates that have successfully accessed the BS in the previous TTI through either the pre-allocation scheme (denoted as $\mathcal{S}_{t-1}$) or the conventional dynamic access scheme (denoted as $\mathcal{C}_{t-1}$). Without doubt, those candidates should belong to $\Psi$. Then the BS performs a dynamic strategy $\Pi^{dynamic}$ based on a sequential learning algorithm dealing with the AMAB problem for each candidate $y\in\Theta_t$, which aims to choose the reserved nodes $\mathbf{k_y}$ with high probability to transmit data in this TTI. Finally, the BS gets all the reserved nodes in $\Omega$ and  implements the pre-allocation of RBs for them.

\begin{figure*}[t]
            \centering
            \includegraphics[height=6.4cm, width=0.81\textwidth]{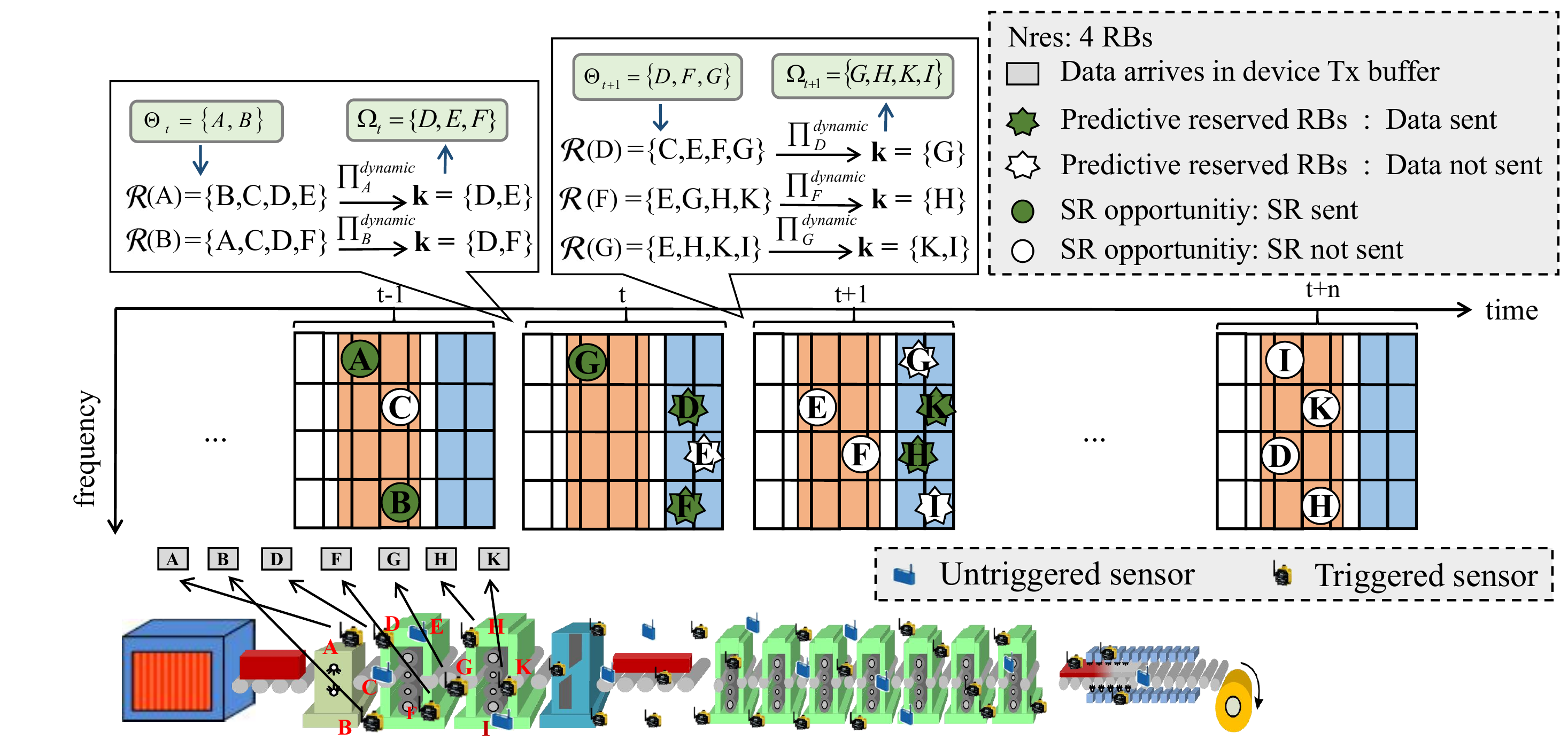}
            \vspace{-3pt}
            \caption{An Example of the Dynamic Stage.}
            \label{AMAB}
            \vspace{-18pt}
    \end{figure*}

\textbf{Example:} We illustrate the dynamic stage using an example in Fig~\ref{AMAB}. At the beginning of TTI $t$ and $t+1$, the BS collects $\Theta$ according to the accessed nodes in the previous TTI and performs $\Pi^{dynamic}$ to select the final reserved node set $\Omega$ with $N_{res}$ = 4 RBs. If the SR opportunity of a node (i.e. the triggered sensor G in this example) arrives before it is notified a reserved RB, the node will be scheduled in the traditional way. Note that through the pre-allocation scheme, the triggered nodes $D,F,K,H$ successfully access the BS before their traditional SR opportunities while the untriggered nodes $E,I$ waste their pre-allocated RBs. Thus, the dynamic strategy $\Pi^{dynamic}$ should learn and explore the most valuable correlated nodes to improve the prediction accuracy. The details will be discussed in Section~\ref{dynamic}.
\section{Static Stage Based on the Naive Bayesian Model}\label{static}
As discussed in the previous section, the main task of the static stage is to decide the reservation candidates as well as their static reservation sets, which should consist of the most correlated nodes. To this end, we propose to explore the correlation between nodes through the access history based on the Naive Bayesian model.
\subsection{Multinomial-event Naive Bayesian Model}
In industrial process automation, the position as well as the sensing type of a triggered node carries valuable information for predicting the nodes that are likely to transmit afterwards. Thus, for each node $x\in\mathcal{N}$, let $L_x$ and $A_x$ denote its location and sensing type respectively. $S_x$ denotes the TTI that the node $x$ has accessed the channel successfully.
In the supervised generative learning, we have a set of $I$ training samples of $\{(\mathbf{x_1},y_1),(\mathbf{x_2},y_2),...,(\mathbf{x_I},y_I)\}$, where an accessed node $y_i \in \mathcal{N}$ is the label of the $i$-th sample, and its feature nodes $\mathbf{x_i}$ denotes a vector consisting of the potential correlated nodes that are either in the vicinity of $y_i$ (within the distance range of $R_r$), or belongs to the same sensing type as $y_i$. Essentially, the feature nodes must be selected from those nodes that have successfully accessed the BS in nearby subframes (within the time range of $R_t$) of $S_{y_i}$. Thus, $\mathbf{x_i}$ can be defined as follows:
        \begin{equation}
         \mathbf{x_i} = \{x\in\mathcal{N}\mid S_x\in[S_{y_i}-R_t,S_{y_i}+R_t], A_x = A_{y_i}\ or\ |L_x - L_{y_i}|\leq R_r \}    
        \end{equation}
        where $|\cdot|$ represents the Euclidean distance between two nodes.

The reason to use the generative learning algorithm is that for each $y$, the correlation between $y$ and its feature node $x_k\in\mathbf{x}$ is needed to be calculated based on $\mathbb{P}(x_k|y)$, which represents the probability that $x_k$ will request scheduling after $y$ is triggered. In this paper, we adopt the multinomial-event Naive Bayesian model due to its desirable property in terms of short time consumption, good accuracy and ease of implementation~\cite{bayes1}. Moreover, in virtue of its simplifying conditional independence assumption, a naive Bayesian classifier is well applicable in our case since we focus on the correlation between $y$ and its feature node $x_k$ instead of the relationship between feature nodes. 

Let $\mathbf{x_i} = \{x^{(i)}_1,x^{(i)}_2,...,x^{(i)}_{n_i}\}$ denotes the $i$-th sample's feature vector. The parameters for the model $\theta$ are given by: \vspace{-3pt}
        \begin{equation}
        \phi_{p|q} = \mathbb{P}(x_k = p|y = q),\ \ \  \phi_{q} =  \mathbb{P}(y = q)   \ \ p,q \in \mathcal{N}
        \end{equation}
Then the joint likelihood can be defined as follows:\vspace{-3pt}
        \begin{equation}\vspace{-3pt}
        \begin{aligned}
         \mathcal{L}(\phi_{p|q},\phi_{q})& = \prod_{i=1}^{I}  \mathbb{P}(\mathbf{x_{i}},y_{i}) \\
         &= \prod_{i=1}^{I} (\prod_{k=1}^{n_i}  \mathbb{P}(x_k^{(i)}|y_i;\phi_{p|q})) \mathbb{P}(y_{i};\phi_{q})
        \end{aligned}
        \vspace{-2pt}
        \end{equation}

In order to maximize the joint likelihood, the model can be trained with the access history, and the Laplace smoothing is adopted to effectively rule out the zero probabilities. Then the model $\theta$ is available for calculating the correlation according to different metrics as will be discussed in the next subsection.
    \subsection{Static Reservation Set Selection}
    In the literature, some metrics have been utilized to measure the correlation between $y$ and its feature node $x_k$, and we will directly give their definitions as follows~\cite{metrics}:
    \begin{enumerate}
    \item \textbf{Posterior probability}: Posterior probability represents the probability that the feature node $x_k$ will request scheduling after $y$ is triggered, which can be simply utilized as a metric to measure the correlation, that is, $\mathcal{E}_{P}(x_k,y)=\mathbb{P}_\theta(x_k|y)$.
    \item \textbf{Mutual information (MI)}: Given the model $\theta$, MI can be used to measure the mutual dependence between two nodes, which is defined as follows:
        \begin{equation}
        \mathcal{E}_{MI}(x_k,y) = \sum_{p\in\{x_k,\sim x_k \}}\sum_{q\in\{y,\sim y\}}\mathbb{P}_\theta(p,q)log_2\frac{\mathbb{P}_\theta(p,q)}{\mathbb{P}_\theta(q)\mathbb{P}_\theta(p)}
        \end{equation}
        where $\mathbb{P}_\theta(p,q)$ is calculated by $\mathbb{P}_\theta(p|q) \ast \mathbb{P}_\theta(q)$.

    \item \textbf{Chi-square test}: A chi-squared test (or $\chi^2$ test) is an efficient way to evaluate whether two nodes are dependent. Specifically, the Chi-square test based correlation metric is defined as:
        \begin{equation}
        \mathcal{E}_{\chi^2}(x_k,y) = \sum_{p\in\{x_k,\sim x_k \} }\sum_{q\in\{y,\sim y\}}\frac{(N_{pq}-E_{pq})^2}{E_{pq}}
        \vspace{-2pt}
        \end{equation}
        where $N_{pq} = I \mathbb{P}_\theta(p,q) $ and $I$ is the total number of samples. $E_{pq}$ represents the expected value of independence, which is defined as follows:
         \begin{equation}
         E_{pq} = I\ast \mathbb{P}_\theta(p)\ast \mathbb{P}_\theta(q) = I\ast \mathbb{P}_\theta(q)\sum_{q=y_1}^{y_I} \mathbb{P}_\theta(p|q)\mathbb{P}_\theta(q)
         \vspace{-3pt}
        \end{equation}
    \end{enumerate}

The difference between $\mathcal{E}_{MI}(x_k,y)$ and $\mathcal{E}_{\chi^2}(x_k,y)$ lies in that those less frequently triggered nodes will have a higher ranking in Chi-square test than in the MI metric. Thus, $\chi^2$-based correlation metric is more appropriate when those highly related nodes are less likely to transmit packets due to network dynamics. However, with small size of selected feature nodes, the accuracy of $\chi^2$ test will become worse than MI due to the interference brought by infrequently triggered nodes.

Two problems remain in the static stage. Firstly, is it necessary to pre-allocate RBs for the correlated nodes of $y$ if it is accessing the BS currently? Secondly, if so, which potential correlated nodes should be selected into its static reservation set $\mathcal{R}(y)$? To this end, we propose a simple threshold-based static strategy $\Pi^{static}$ to select the reservation candidates:
     \begin{equation}
      \Pi^{statuc}=
      \begin{cases}
       1, & ~\mbox{if}\ \  max\ \mathcal{E}(x_k,y) \geq \alpha  \\
       0, & ~\mbox{otherwise}
       \end{cases}
       \vspace{-3pt}
      \end{equation}

In this strategy, if the maximum $\mathcal{E}(x_k,y)$ is below a prescribed threshold $\alpha$, then it means that none of $y$'s feature nodes has a strong correlation with it. Thus, it is worthless to reserve RBs according to $y$. Otherwise, $y$ should serve as a reservation candidate in $\Psi$, and a static reservation set $\mathcal{R}(y)$ with a fixed size of $\xi$ will be selected for it as follows:
        \begin{equation}
         \mathcal{R}(y) = max_\xi\ \mathcal{E}(x_k,y)
         \vspace{-3pt}
        \end{equation}
        where $max_\xi$ denotes the $\xi$ most correlated nodes in the feature nodes of $y$. The value of $\xi$ should not only ensure that $ \mathcal{R}$ contains enough correlated nodes, but also consider the complexity in the dynamic stage.
     \vspace{-3pt}
\section{Dynamic Stage Based on the Sequential Learning}~\label{dynamic}
As mentioned in Section~\ref{overview}, the reasons for the dynamic stage are twofold. Firstly, although many un-correlated nodes have been ruled out of the reservation set in the static stage, not all the nodes in $\mathcal{R}(y)$ will transmit after $y$ accessing the BS due to intermittent network dynamics. Moreover, it is difficult to pre-allocate for all the nodes in the static reservation sets of so many candidates within the limited reserved RBs.
Secondly, the packet deadlines of the nodes in $\mathcal{R}(y)$ vary. Despite successfully being predicted and reserved, the packet will be outdated if its latency has already exceeded the prescribed deadline, which is a waste of reserved RBs.
 Therefore, the objective of the dynamic stage is to maximize the rewards of the reserved RBs, which involves a fundamental trade-off between exploration (learning the trigger statistics of nodes in the static reservation set) and exploitation (reserving for nodes with the best payoff). To this end, we propose a dynamic reservation prediction (DRP) strategy $\Pi^{dynamic}$ based on EXP3~\cite{EXP3}, which stands the exponential-weight algorithm for exploration and exploitation dealing with the AMAB problem.
      \vspace{-3pt}
\subsection{Weighted Reserved Resources Allocation}
Due to the limited RBs, the first step of the dynamic stage is to decide how many RBs should be shared by each reservation candidate. To improve the prediction accuracy, it is desirable to allocate more reserved RBs to the candidates with stronger correlation with their selected feature nodes in $\mathcal{R}$. Formally, the number of RBs $\delta_{y}$ that can be allocated to a candidate $y \in \Theta_t $ is a weighted share of $N_{res}$ as follows:
        \begin{equation}
         \delta_{y} = \frac{\sum_{x\in \mathcal{R}(y)}\mathcal{E}(x,y) }{ \sum_{q\in \Theta_t }\sum_{p\in \mathcal{R}(q)} \mathcal{E}(p,q)} N_{res}
        \end{equation}
        where $\mathcal{E}(x,y)$ can be any of the three correlation metrics introduced in the previous section.

Note that $\delta_{y}$ corresponds to the number of nodes can be selected from $\mathcal{R}(y)$ to reserve for. Here we only fucus on the number of reserved RBs, while the scheduling problem for each reserved RB is not involved in this paper.
        \vspace{-3pt}
\subsection{Dynamic Reservation Prediction Strategy}
In the following, we focus on the dynamic strategy $\Pi^{dynamic}$ (DRP) based on EXP3 for each candidate in $\Omega$. For a candidate $y$, the BS determines which $\delta_{y}$ selected feature nodes in $\mathcal{R}(y)$ should be pre-allocated for in the current TTI. Here, each choice represents an arm for $y$, which can be defined as follows: $\mathbf{k} = (x_1^{\mathbf{k}},x_2^\mathbf{k},...,x_{\delta_{y}}^\mathbf{k})$ where $ x \in \mathcal{R}(y) $. Let $\mathbb{K}_y$ denotes the arm space, which consists of all possible assignments and $\mathcal{K} = |\mathbb{K}_y|$ denotes the number of available arms.

Before devoting into the details of DRP, we firstly define the sigmoidal utility for the reward of a successfully reserved node $x$ as in~\cite{utilitys}, which is a two-tuples $(a_x, b_x)$ function of the specific latency request, where $a_x$ and $b_x$ denote the criticality and the nominal delay threshold (inflection point of~(\ref{utility})) of the sensor type $A_x$ of $x$, respectively. All nodes belonging to the same sensor type have the same sigmoidal utility as follows:
     \begin{equation}\label{utility}\vspace{-8pt}
         U_{A_x}(l_x^t) = 1-c_x(\frac{1}{1+e^{-a_x(l_x^t-b_x)}}-d_x)
     \end{equation}
     where $$ c_x=\frac{1+e^{a_xb_x}}{e^{a_xb_x}}, d_x=\frac{1}{1+e^{a_xb_x}}$$ and $l_x^t$ represents the latency when $x$ successfully accesses the BS at TTI $t$. This does not decrease much until the latency exceeds the delay threshold (i.e., $l_x^t > b_x$), which motivates the BS to pre-allocate for packets within their deadlines.

    \begin{algorithm} [t]
    \caption{Dynamic Reservation Prediction Strategy}
    \label{A3}
    \KwIn{$\mathbb{K}_y,\mathcal{R}(y)$}
    \KwOut{The selected reserved arm $\mathbf{k_s}$ in each trial $s$}
    \textbf{Initialization:} $\omega_\mathbf{j}(1) =1$ for all $\mathbf{j} \in \mathbb{K}_y$ ; $\gamma \in (0,1]$\\
    \For{$s = 1,2,3...$}
    {   1. Set the probability for each arm $\mathbf{j}\in \mathbb{K}_y$:
        \begin{equation} \label{Parm}
        P_{\mathbf{j},s} = (1-\gamma)\frac{\omega_\mathbf{j}(s)}{\sum_{\mathbf{k} = 1}^ {\mathcal{K}}\omega_{\mathbf{k}}(s)} + \gamma p(\mathbf{j};\theta)
        \end{equation}
        \ \ \ where
        \begin{equation}
        p(\mathbf{j};\theta) =\frac{ \prod_{x_j \in \mathbf{j}} \mathbb{P}_\theta(x_j|y) }{\sum_{\mathbf{k}\in \mathbb{K}_y} \prod_{x_k \in \mathbf{k}} \mathbb{P}_\theta(x_k |y) }
        \end{equation}
       2. Select $\mathbf{k_s}$ according to $\mathbf{P_s} = \{P_{\mathbf{1},s},P_{\mathbf{2},s},...,P_{\mathbf{\mathcal{K}},s}\}$ \;
       3. Calculate $\mathbf{k_s}$'s reward $\in[0,1]$ based on the observed feedback:
        \begin{equation}\label{lalal}
        r_{\mathbf{k_s},s} = (\sum_{x_{k} \in \mathbf{k_s}} U_{A_{x_k}}(l_{x_k}^{t_s}) \cdot \mathbb{I}_{\{x_k \in \mathcal{S}_{t_s}\}} - \beta\sum_{x_{k} \in \mathbf{k_s}}\mathbb{I}_{\{x_k \notin \mathcal{S}_{t_s}\}}) /\delta_{y}
        \end{equation}
       4. Estimate the rewards $\in[0,1]$ of other arms $\mathbf{j}$:
        \begin{equation}
        r_{\mathbf{j},s} = min\{(\sum_{x_j \in \mathbf{j} }  U_{A_{x_j}}(l_{x_j}^{t_s})\cdot \mathbb{I}_{\{x_j \in \mathcal{S}_{t_s}\}}  - \beta\sum_{x_j \in \mathbf{j}} \mathbb{I}_{\{x_j \in \mathcal{F}_{t_s}\}}) / \delta_{y},  r_{\mathbf{k_s},s} \}
        \end{equation}
        \ \ \ \ where \ \ \ \ \ \ \ \ \ $\mathcal{F}_s = \{x_k \in \mathbf{k_s} | x_k \notin \mathcal{S}_s  \}$  \\
       5. Update the weight of each arm : \\
       \For{$\mathbf{k} = \mathbf{1},...,\mathcal{K}$}{
         \begin{equation}
         \hat{r}_{\mathbf{k},s} =
         \begin{cases}
          r_{\mathbf{k},s}/P_{\mathbf{k},s} & \mathbf{k} = \mathbf{k_s}  \\
          r_{\mathbf{k},s}/{max \{P_{\mathbf{k},s}, 1-P_{\mathbf{k},s} \}} & \mathbf{k} \neq \mathbf{k_s}
         \end{cases}
         \end{equation}
          \begin{equation}
           \omega_\mathbf{k}(s+1) = \omega_\mathbf{k}(s) exp(\frac{\gamma\hat{r}_{\mathbf{k},s}}{\mathcal{K}}) \ \ \ \
           \vspace{-8pt}
         \end{equation}
       }
     }
     \vspace{-3pt}
    \end{algorithm}
     The DRP strategy for a reservation candidate $y$ is summarized in Algorithm~\ref{A3}. Each learning trial $s$ starts when the BS receives the transmission from the node $y$. $\omega_\mathbf{k}(s)$ denotes the weight of the arm $\mathbf{k}$ at $s$, which is updated at each trial. In the trial $s$, we firstly calculate the probability distribution of all arms $\mathbf{P_s} = \{P_{\mathbf{1},s},P_{\mathbf{2},s},...,P_{\mathbf{\mathcal{K}},s}\}$  and decide the reserved arm $\mathbf{k}_s$ based on it. Then, all the reserved arms selected through $\Pi^{dynamic}$ for each candidate in $\Theta_t$ are incorporated into the final reserved nodes set $\Omega$, and the reserved RBs are pre-allocated to them accordingly. After this trial, the successful pre-allocated node set $\mathcal{S}_{t_s}$ can be observed, where $t_s$ denotes the TTI of the trial $s$. Then the reward of $\mathbf{k}_s$ is calculated and other arms' rewards are estimated accordingly. Note that the publishing terms in (13) and (14) are achieved through different sets, where $\mathcal{F}_s$ is the set containing those wrongly predicted nodes in $\mathbf{k}_s$. The reason is that with the partial information obtained from this feedback, it is impossible to draw the conclusion that other nodes in $\mathbf{j}$ (not involved in $\mathbf{k}_s$) will fail to access the BS in the reserved RBs at $t_s$.

    Although the exploration occurs only when $y$ is triggered, the convergence and accuracy of DRP can be improved from two aspects. First, we partially explore each arm according to the learning model in the static stage rather than uniform search in the exploring part of (11). This works in two situations: 1) when the correlation between $y$ and its selected feature nodes in $\mathcal{R}(y)$ are diverse, which means the variance of those correlation values is large, then the BS is more likely to pre-allocate for the arm that involves nodes with larger triggered probability; 2) when the variance is small, the exploring term is similar with uniform distribution and will have more chance to explore each arm. Second, the rewards of other arms are estimated even they are not chosen in step 2 in order to reduce the exploration.

    Another benefit of Algorithm~\ref{A3} is the customized utility functions for different sensing applications, with which the packets received in the reserved RBs are less likely to become outdated. So the reserved RBs will receive more rewards in Step 3 and 4, where $l_{x}^{t_s}$ denotes the latency of the packet delivered by $x$ when it arrives at the BS in TTI $t_s$.
\subsection{Regret Bound of DRP}
In this following, we analyze the performance of DRP in terms of the regret bound under the most dynamic case, where the BS explores arms with uniform distribution. This holds for each candidate performing the DRP strategy $\Pi^{dynamic}$.

    \emph{Theorem 1:} For any $\mathcal{K}>0$ and  any $\gamma \in (0,1]$,
    \begin{equation}~\label{D0}
    \mathbb{R}_{DRP} \leq \frac{1- \gamma}{\gamma} ln\mathcal{K} + \frac{\gamma (2e-3) + \mathcal{K}-1}{\mathcal{K}}G_{max}
    \end{equation}
    holds for any assignment of rewards and any $S>0$. $\mathbb{R}_{DRP} = G_{max}- \mathbf{E}[G_{DRP}]$, in which $G_{max}$ and $G_{DRP}$ denote the gains of the single globally best action at trail horizon S and actions chosen by DRP, respectively.

    Comparing with the regret of EXP3~\cite{EXP3}, that is,
    \vspace{-1pt}
    \begin{equation}
    \mathbb{R}_{EXP3} \leq \frac{\mathcal{K}ln\mathcal{K}}{\gamma} +(e-1)\gamma G_{max},
    \vspace{-1pt}
    \end{equation}
    it is easy to show that for all $\mathcal{K}>0 $, $\mathbb{R}_{DRP} < \mathbb{R}_{EXP3}$.

    Since $\mathbb{R}_{EXP3}$ can be further bounded by $2.63\sqrt{g\mathcal{K}ln\mathcal{K}}$ with the input parameter: \vspace{-3pt} $$\gamma = min \{1,\sqrt{\frac{\mathcal{K}ln\mathcal{K}}{(e-1)g}}\},$$ where $g$ is the upper bound of $G_{max}$, shown in Corollary 3.2~\cite{EXP3}, it suffices to show that the regret of DRP also scales with $\sqrt{g}$, which means that the regret increases with $\sqrt{S}$ as the reward is no more than 1.

    The main improvement of $\mathbb{R}_{DRP}$ lies in the way it scales with the number of arms $\mathcal{K}$. To understand this, it is indeed quite pertinent to show how the bound increases with $\mathcal{K}$.

    \emph{Corollary 1.1:} For any $S>0$, assume that $g\geq G_{max}$ and the DRP algorithm is run with the input parameter
    $$\gamma = min \{1,\sqrt{\frac{\mathcal{K}ln\mathcal{K}}{(2e-3)g}}\},$$
    then
    \vspace{-5pt}
    \begin{equation}
    \mathbb{R}_{DRP} \leq 2\sqrt{\frac{(2e-3)ln\mathcal{K}g}{\mathcal{K}}} + g \leq 3.12 \sqrt{\frac{ln\mathcal{K}g}{\mathcal{K}}} +g
    \end{equation}
    holds for any assignment of rewards.

        \begin{figure*}[t]
        \centering
        \setlength{\abovecaptionskip}{0.5cm}
        \begin{minipage}[t]{0.323\linewidth}
            \subfigure[The reservation set size comparison ($N_{res} = 50$)]{
            \includegraphics[width=1\textwidth,height=4.8cm,]{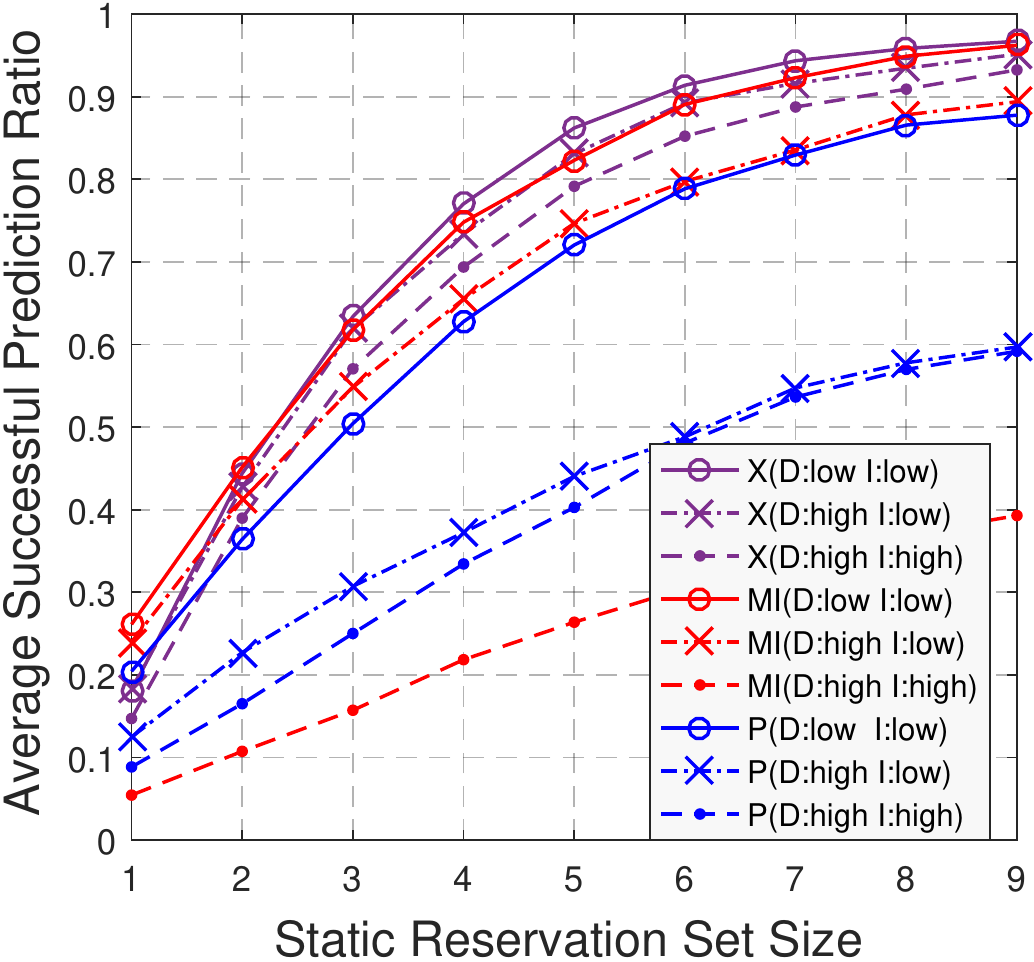}}
         \end{minipage}
        \begin{minipage}[t]{0.323\linewidth}
            \subfigure[The reserved RBs number comparison ($\xi = 8$)]{
            \includegraphics[width=1\textwidth,height=4.8cm,]{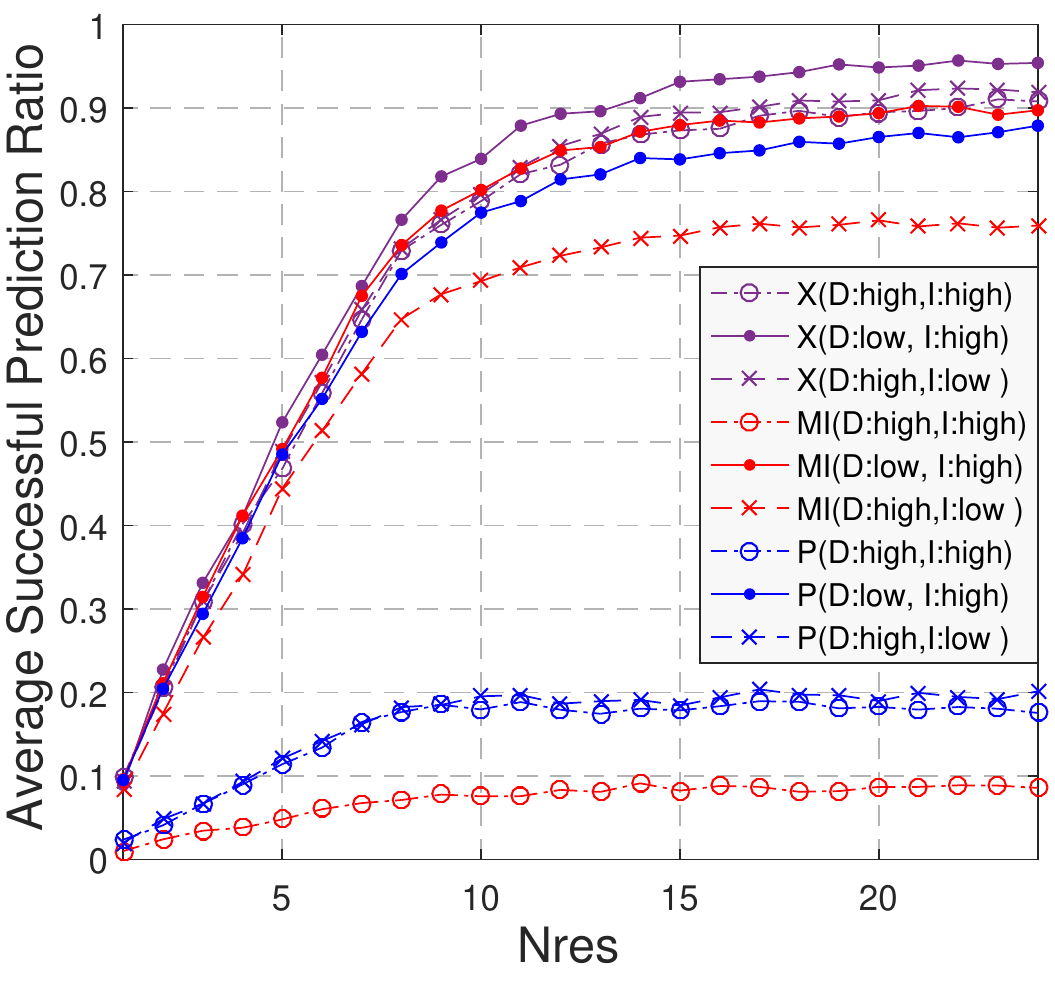}}
         \end{minipage}
         \begin{minipage}[t]{0.323\linewidth}
            \subfigure[The threshold comparison ($I:high, D:high$)]{
            \includegraphics[width=1\textwidth,height=4.8cm,]{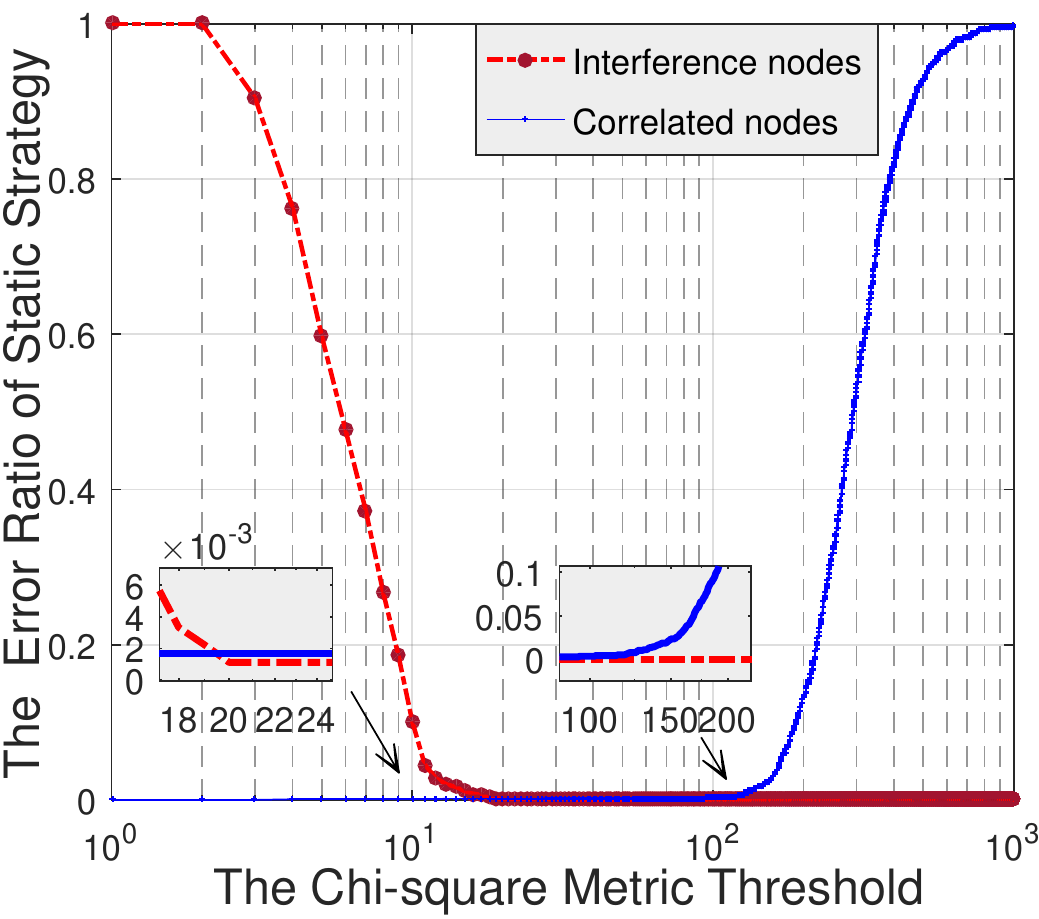}}
         \end{minipage}
         \vspace{-13pt}
        \caption{The performance comparison of three correlation metrics with different static parameters.}
        \label{simulation1}
        \vspace{-8pt}
        \end{figure*}
                \begin{figure*}[t]
        \centering
        \setlength{\abovecaptionskip}{0.5cm}
        \begin{minipage}[t]{0.323\linewidth}
            \subfigure[The performance with low dynamics. ($N_{res} = 6, \xi=8, \alpha = 50$)]{
            \includegraphics[width=1\textwidth,height=4.8cm,]{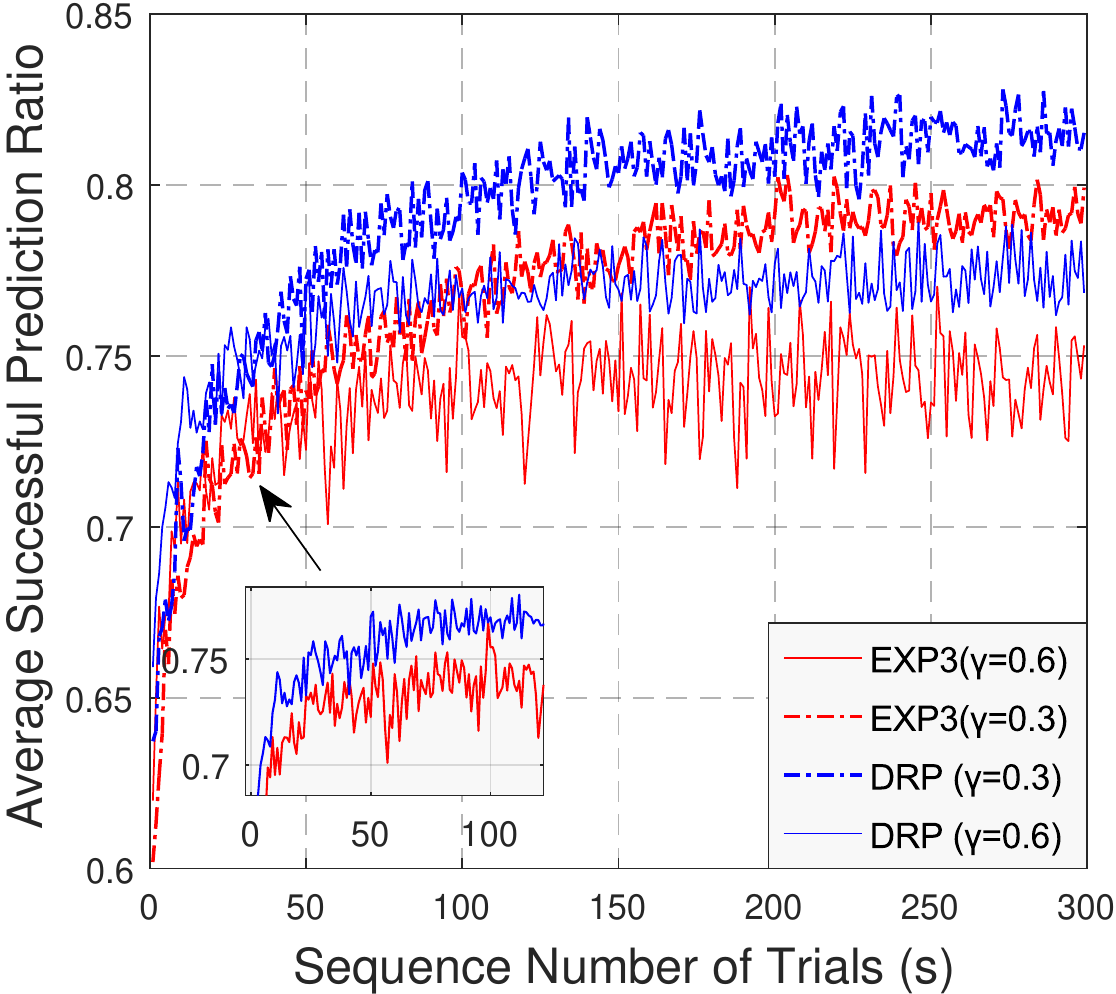}}
         \end{minipage}
        \begin{minipage}[t]{0.323\linewidth}
            \subfigure[The performance with high dynamics. ($N_{res} = 6, \xi=8, \alpha = 50$)]{
            \includegraphics[width=1\textwidth,height=4.8cm,]{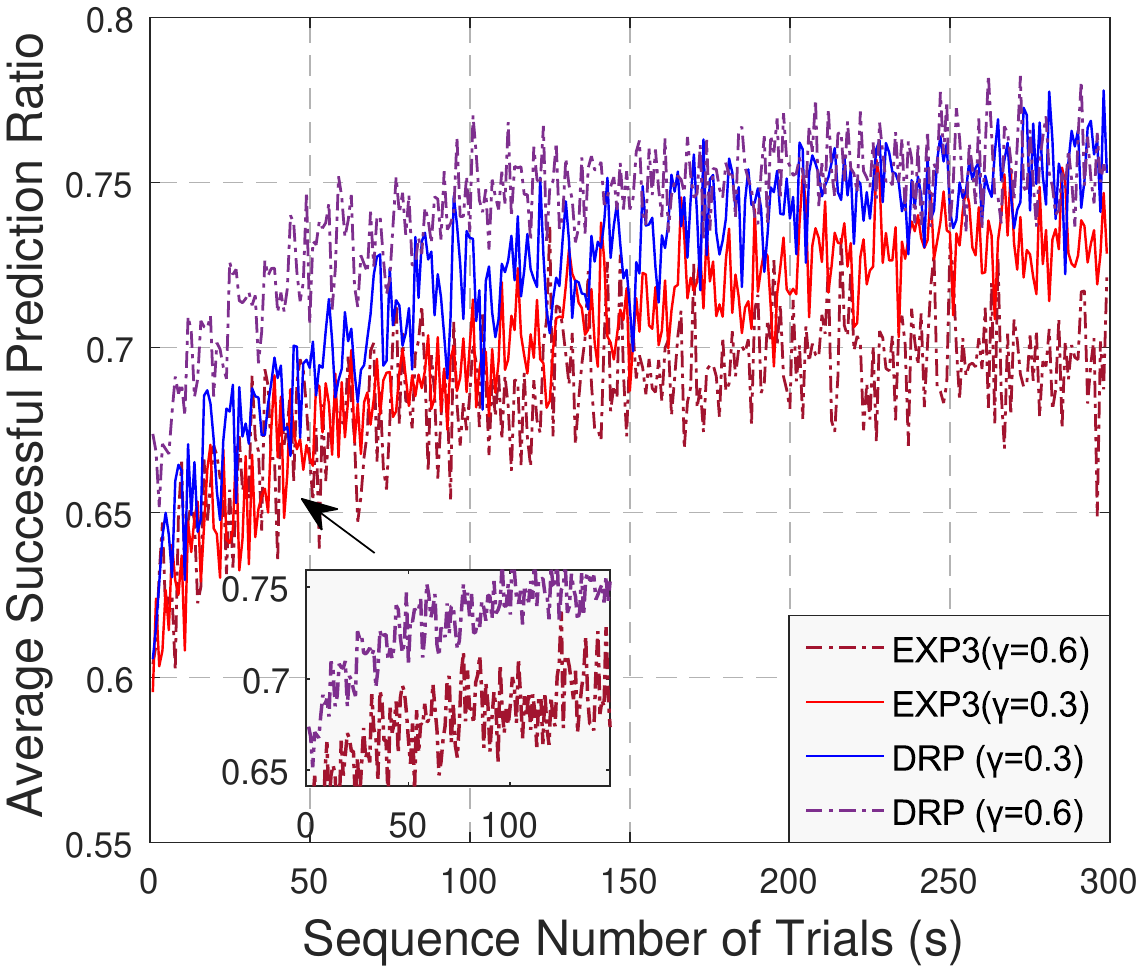}}
         \end{minipage}
         \begin{minipage}[t]{0.323\linewidth}
            \subfigure[The performance with different utilities.($N_{res} = 6, \xi=8$, $\gamma = 0.3, \alpha = 50$, D = low)]{
            \includegraphics[width=1\textwidth,height=4.8cm,]{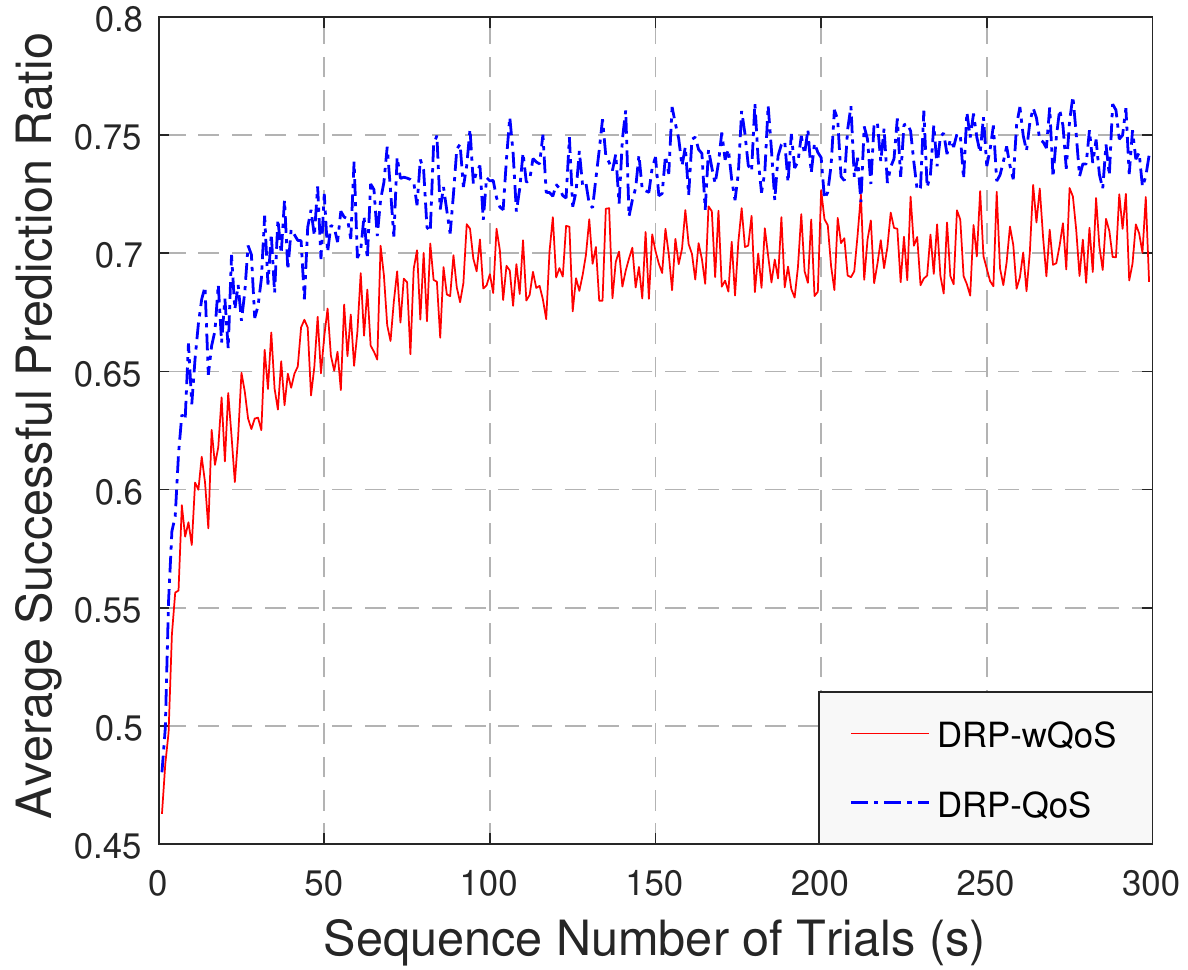}}
         \end{minipage}
         \vspace{-13pt}
        \caption{The performance of DRP with different dynamics and temporal reward functions.}
        \label{simulation2}
        \vspace{-18pt}
        \end{figure*}
    \emph{Proof of Corollary 1.1:} If $g\leq \mathcal{K}ln\mathcal{K}/(2e-3)$, then $\gamma = 1$ and thus the bound is trivial since the expected regret cannot be more than $((2e-3)+\mathcal{K} -1)g/\mathcal{K}$. Otherwise, $\gamma = \sqrt{\mathcal{K}ln\mathcal{K}/(2e-3)g}$. By Theorem 1, the expected regret is at most
    \begin{equation}
    \begin{aligned}
    &\frac{1- \gamma}{\gamma} ln\mathcal{K} + \frac{\gamma (2e-3) + \mathcal{K}-1}{\mathcal{K}}\mathcal{G}_{max}\\
    &\leq \frac{1}{\gamma} ln\mathcal{K} + \frac{\gamma (2e-3)}{\mathcal{K}}\mathcal{G}_{max}  + \mathcal{G}_{max} \leq 2\sqrt{\frac{(2e-3)ln\mathcal{K}g}{\mathcal{K}}} + g
    \end{aligned}
    \end{equation}
    as desired. $\ \ \ \ \ \ \ \ \ \ \ \ \ \ \ \ \ \ \ \ \ \ \ \ \ \ \ \ \ \ \ \ \ \ \ \ \ \ \ \ \ \ \ \ \ \ \ \ \ \ \ \ \ \ \ \blacksquare$

    Obviously, the bound of DRP scales better as a function of $\sqrt{ln\mathcal{K}/\mathcal{K}}$ rather than $\sqrt{\mathcal{K}ln\mathcal{K}}$,  which is incomparable to our bound when $\mathcal{K}$ is large (e.g., $\xi$ should be large enough so that $\mathcal{R}$ involves sufficient correlated nodes at the expense of an acceptable increase complexity).

    \emph{Proof of Theorem 1:} The proof follows the procedure of Theorem 3.1 in~\cite{EXP3} with some modifications. Here we use the following simple facts, which are the main differences between the bound provided in this theorem and the previous result, which can be immediately derived from the definitions:\vspace{-3pt}
    \begin{equation}~\label{D1}
    \begin{aligned}
         &\sum_{\mathbf{k}=1}^{\mathcal{K}} P_{\mathbf{k},s}\hat{r}_{\mathbf{k},s}  \\
         &= P_{\mathbf{k_s},s}\frac{r_{\mathbf{k_s},s}}{P_{\mathbf{k_s},s}} + \sum_{\mathbf{k} \neq \mathbf{k_s} }  P_{\mathbf{k},s} \frac{r_{\mathbf{k},s}}{max\{P_{\mathbf{k},s}, 1-P_{\mathbf{k},s}\}} \\
         &\leq r_{\mathbf{k_s},s} + (\mathcal{K} -1) r_{\mathbf{k_s},s} \leq \mathcal{K} r_{\mathbf{k_s},s}
    \end{aligned}\vspace{-5pt}
    \end{equation}
    \begin{equation}~\label{D2}\vspace{-3pt}
    \begin{aligned}
    &\sum_{\mathbf{k}=1}^{\mathcal{K}}  P_{\mathbf{k},s}\hat{r}_{\mathbf{k},s}^2 \leq  r_{\mathbf{k_s},s}  \hat{r}_{\mathbf{k_s},s} + \sum_{\mathbf{k} \neq \mathbf{k_s} } r_{\mathbf{k_s},s} \hat{r}_{\mathbf{k},s} \leq \sum_{\mathbf{k}=1}^{\mathcal{K}} \hat{r}_{\mathbf{k},s}
    \end{aligned}
    \end{equation}
    \begin{equation}~\label{smaller}
         \hat{r}_{\mathbf{k},s}\leq 1/max\{P_{\mathbf{k},s}, 1-P_{\mathbf{k},s}\} \leq 1/P_{\mathbf{k},s} \leq \mathcal{K}/\gamma
    \end{equation}

    Let $W_s = \omega_\mathbf{1}(s) +\cdots+ \omega_\mathcal{K}(s)$ denote the sum of weights. For all sequences $\mathbf{k_1},...,\mathbf{k_S}$ of actions drawn by our dynamic reservation prediction strategy, we have:
    \begin{equation}
    \begin{aligned}
     \frac{W_{s+1}}{W_s}& = \sum_{\mathbf{k}=1}^\mathcal{K} \frac{\omega_\mathbf{k}(s+1)}{W_s} = \sum_{\mathbf{k}=1}^\mathcal{K} \frac{\omega_\mathbf{k}(s)}{W_s} exp(\frac{\gamma}{\mathcal{K}} \hat{r}_{\mathbf{k},s}) \\
     (24.a)\ \ \ & = \sum_{\mathbf{k}=1}^\mathcal{K} \frac{P_{\mathbf{k},s}-\gamma p(\mathbf{k};\theta)}{1-\gamma} exp(\frac{\gamma}{\mathcal{K}} \hat{r}_{\mathbf{k},s}) \\
     (24.b)\ \ \ & \leq \sum_{\mathbf{k}=1}^\mathcal{K} \frac{P_{\mathbf{k},s}-\gamma p(\mathbf{k};\theta)}{1-\gamma}[1 + \frac{\gamma}{\mathcal{K}} \hat{r}_{\mathbf{k},s} +(e-2)(\frac{\gamma}{\mathcal{K}}  \hat{r}_{\mathbf{k},s})^2 ]\\
     (24.c)\ \ \ & \leq 1 +   \frac{\gamma}{(1-\gamma)\mathcal{K}} \sum_{\mathbf{k}=1}^\mathcal{K}P_{\mathbf{k},s} \hat{r}_{\mathbf{k},s} + \frac{e-2}{1-\gamma}(\frac{\gamma}{\mathcal{K}})^2 \sum_{\mathbf{k}=1}^\mathcal{K}P_{\mathbf{k},s} \hat{r}_{\mathbf{k},s} ^2 \\
     (24.d)\ \ \ & \leq 1 + \frac{\gamma}{(1-\gamma)}r_{\mathbf{k_s},s} + \frac{e-2}{1-\gamma}(\frac{\gamma}{\mathcal{K}})^2  \sum_{\mathbf{k}=1}^{\mathcal{K}} \hat{r}_{\mathbf{k},s}
    \end{aligned}
    \end{equation}
    where (24.a) uses the definition of $P_{\mathbf{k},s}$. (24.b) uses~(\ref{smaller}) and the fact that $e^x \leq 1+x+(e-2)x^2 $ for $x\leq 1$ ; (24.d) uses~(\ref{D1}) (\ref{D2}). Taking logarithms and using $1+x\leq e^x$ yields
     \vspace{-3pt}
    \begin{equation}
     ln \frac{W_{s+1}}{W_s} \leq \frac{\gamma}{(1-\gamma)} r_{\mathbf{k_s},s} + \frac{e-2}{1-\gamma}(\frac{\gamma}{\mathcal{K}})^2  \sum_{\mathbf{k}=1}^{\mathcal{K}} \hat{r}_{\mathbf{k},s}
    \end{equation}
    Summing over s, we then obtain:
    \begin{equation}\label{D3}
    ln \frac{W_{S+1}}{W_1} \leq \frac{\gamma}{(1-\gamma)} G_{DRP} + \frac{e-2}{1-\gamma}(\frac{\gamma}{\mathcal{K}})^2 \sum_{s=1}^S \sum_{\mathbf{k}=1}^{\mathcal{K}}\hat{r}_{\mathbf{k},s}
    \end{equation}
    For any action $\mathbf{j}$,
    \begin{equation}
    ln \frac{W_{S+1}}{W_1} \geq ln \frac{\omega_\mathbf{j}(S+1)}{W_1} = \frac{\gamma}{\mathcal{K}} \sum_{s=1}^S \hat{r}_{\mathbf{j},s} - ln\mathcal{K}
    \end{equation}
    Combining this with~(\ref{D3}), we obtain
    \begin{equation}\label{D4}
    G_{DRP} \geq \frac{1-\gamma}{\mathcal{K}} \sum_{s=1}^S \hat{r}_{\mathbf{j},s} - \frac{1- \gamma}{\gamma} ln\mathcal{K} - \frac{\gamma (e-2)}{\mathcal{K}^2} \sum_{s=1}^S \sum_{\mathbf{k}=1}^{\mathcal{K}} \hat{r}_{\mathbf{k},s}
    \end{equation}

    Taking the expectation of both sides of~(\ref{D4}) with respect to the distribution of
    $\langle \mathbf{k_1},...,\mathbf{k_S} \rangle$, then for the expected value of each $\hat{r}_{\mathbf{k},s}$, we have
    \begin{equation}\label{D5}
    \begin{aligned}
    &r_{\mathbf{k},s} \leq\mathbf{E}[ \hat{r}_{\mathbf{k},s}|\mathbf{k_1},...,\mathbf{k_S}] \\
    & = \mathbf{E}[P_{\mathbf{k},s}\frac{r_{\mathbf{k},s}}{P_{\mathbf{k},s}} + (1-P_{\mathbf{k},s})\frac{r_{\mathbf{k},s}}{max\{P_{\mathbf{k},s}, 1-P_{\mathbf{k},s}\}}] \leq 2 r_{\mathbf{k},s}
    \end{aligned}
    \end{equation}
    Combining~(\ref{D4}) and (\ref{D5}), we have
    \begin{equation}
    \begin{aligned}
    &\mathbf{E}[G_{DRP}] \geq  \frac{1-\gamma}{\mathcal{K}} \sum_{s=1}^S r_{\mathbf{j},s}- \frac{1- \gamma}{\gamma} ln\mathcal{K}- \frac{2\gamma (e-2)}{\mathcal{K}^2} \sum_{s=1}^S \sum_{\mathbf{k}=1}^{\mathcal{K}} r_{\mathbf{k},s}
    \end{aligned}
    \end{equation}
    Since $\mathbf{j}$ is chosen arbitrarily, we have:
    \begin{equation}
    \sum_{s=1}^S r_{\mathbf{j},s} \leq  G_{max}\ \  and \ \ \sum_{s=1}^S \sum_{\mathbf{k}=1}^{\mathcal{K}} r_{\mathbf{k},s} \leq \mathcal{K} G_{max}
    \end{equation}

    Therefore, the regret of our DRP strategy is bounded as follows:
    \begin{equation}
    \mathbb{R}_{DRP} \leq \frac{1- \gamma}{\gamma} ln\mathcal{K} + \frac{\gamma (2e-3) + \mathcal{K}-1}{\mathcal{K}}G_{max}
    \end{equation}

    This bound holds for any assignments and $\mathcal{K} > 0$, which completes the proof.      $\ \ \ \ \ \ \ \ \ \ \ \ \ \ \ \ \ \ \ \ \ \ \ \ \ \ \ \ \ \ \ \ \ \ \ \ \ \ \ \ \ \ \ \ \blacksquare$
\section{Performance Evaluation}\label{performanceevaluation}
    \subsection{Simulation Setup}
In this section, we evaluate the performance of the proposed DPre framework based on some delay-sensitive process monitoring applications deployed along a fixed-sequence steel rolling production line. The temperature, pressure and humidity sensors are randomly distributed in each procedure cell to control manufacture parameters through feedback, and the vibration sensors are deployed uniformly along the production line to diagnose the health of rollers. Those event-triggered sensors are only activated when the steels arrive.
Table~\ref{mytab1} shows the two-tuples utility requirements of different sensing applications. Here, interference denotes the sensors who are distributed and triggered randomly without relationship with the arrivals of steels. We consider two levels for interference and dynamics factors. For simplicity, $I$ denotes the triggered probability of the interference nodes, which can be divided into two groups (high: 0.8, low: 0.4),  while the triggered probability of other sensors represents the network dynamics (D) (high: 0.6-0.8, low: 0.2-0.4). As for latency, packets transmitted through the conventional dynamic access scheme will be delayed for 10-25ms~\cite{5Glow} randomly to access the BS.
    \subsection{Simulation Results}
    The effectiveness of the proposed framework is assessed in terms of the prediction accuracy (successful prediction ratio). Firstly, we focus on the static stage ($R_r = 25ms$, $R_d = 0.5m$ for selecting samples). The performances of three correlation metrics (X: Chi-square test, MI: Mutual information, P: Posterior probability) are compared. Then, the performance of DRP is compared with the original EXP3 algorithm. Finally, we evaluate the effectiveness of the whole framework.

    \subsubsection{Static Stage Results}
    Fig~\ref{simulation1}(a)(b) show how the performance of three correlation metrics improves with the increasing number of selected feature nodes and reserved resources. It can be concluded that the $\chi^2$ metric provides the best prediction accuracy and resistance facing with high dynamics and interference nodes since it pays more attention to those less frequently triggered correlated nodes, but its accuracy is worse than MI when $\xi$ is small. The second is the MI metric, which however can be easily confused by those frequently triggered interference nodes. Although the posterior probability metric shows a slightly better noise-resistant performance than MI, it is sensitive to network dynamics. Moreover, to ensure $\mathcal{R}$ involves correlated nodes as many as possible, we set $\xi=8$ in the next simulations considering the complexity of the dynamic stage. However, it is unrealistic that the reserved resources are sufficient ($N_{res} = 50$ in Fig~\ref{simulation1}(a)) to accommodate all the correlated nodes. Assuming that the BS just pre-allocates 6 reserved RBs to this production line, it can be seen from the curve of X(D:high,I:high) in Fig~\ref{simulation1}(b) that the original accuracy is about 0.55, which is very low. Therefore, it is necessary to further explore through the DRP algorithm.

    To show the impact of threshold $\alpha$ on the probability of wrongly pre-allocating for interference nodes or refusing to make a reservation for correlated nodes, Fig~\ref{simulation1}(c) delineates the error ratio of $\Pi^{static}$ with increasing $\alpha$. Here, we adopt the $\chi^2$ metric as an example. It can be seen that the curve of interference nodes descends more steeply than the rising trend of correlated nodes and the thresholds in [20,140] can separate these two parts easily. This result confirms that the threshold-based static strategy can not only rule out noise nodes but also prevent omitting correlated nodes by mistake. The same result can also be achieved with the other two metrics.
    \vspace{-0pt}
        \begin{table}[t]
    \centering
        \caption{\textsc{Simulation Settings}}\label{mytab1}
        \begin{tabular}{|m{18pt}<{\centering}|m{34pt}<{\centering}|m{33pt}<{\centering}|m{30pt}<{\centering}|m{30pt}<{\centering}|m{32pt}<{\centering}|}
            \hline
                  & Temperature & Humidity & Pressure & Vibration & Interference \\
            \hline
            $\sharp$of Node & 120 & 120 & 120 & 100 & 300 \\
            \hline
             QoS  & (8ms,0.8) & (12ms,0.45) & (16ms,0.4) & (10ms,0.6) & - \\
            \hline
        \end{tabular}
        \label{tabel}
        \vspace{-14pt}
    \end{table}
    \subsubsection{Dynamic Stage Results} 
    The prediction accuracy of DRP and EXP3 is compared in Fig~\ref{simulation2}(a)(b) with different dynamics. Here, a trial refers to the process that one steel plate goes through the whole production line. Note that with larger $\gamma$, the convergence rate will be faster due to larger weight update rate but at the cost of opportunities to explore more arms. Therefore, with low dynamics, both algorithms with $\gamma = 0.6$ keep a relatively low level with larger jitter comparing with $\gamma=0.3$ since they overweigh randomly chosen arms.
    However, DRP has better prediction accuracy and converges faster than EXP3 regardless of $\gamma$ under both high and low dynamic conditions due to the extra correlation information provided by the static stage, as well as the partial information utilized in unselected arms estimation.

    In Fig~\ref{simulation2}(b), both algorithms have a lower prediction accuracy and convergence rate as well as a larger fluctuation due to high dynamics. Nevertheless, DPR can achieve much better performance with $\gamma = 0.6$, which even exceeds that with $\gamma = 0.3$. The rationale is that under varying environment, the weights of arms are no longer accurate. DRP can pay more attention to the exploration phase, where nodes with different dynamic levels can be explored differentially, while EXP3 still explores all nodes with uniform probability. Therefore, DRP can adaptively adjusts its focus on exploring process to adapt to the network dynamics. Another benefit of DRP can be observed in Fig~\ref{simulation2}(a)(b) that with $\gamma = 0.6$, DRP performs much better than EXP3 in the first few trials (the highest achievable is 0.68 in Fig~\ref{simulation2}(b)) because DRP is able to exploit the static information while EXP3 has nothing in the beginning.

    Taking the delay of nodes into account, the successfully predicted nodes should be the nodes that deliver packets not only in their reserved RBs but also within deadlines. Fig~\ref{simulation2}(c) compares the prediction accuracy of DRP with different reward functions. DRP-wQoS denotes that the sigmoidal utilities of successfully predicted nodes are constant (equal to  1) instead of the decreasing functions (seen in Table~\ref{tabel}) in DRP-QoS. Therefore, both algorithms start with a lower accuracy compared with Fig~\ref{simulation2}(a) as nodes whose access latency has already exceeded thresholds are excluded out of the successful predicted nodes set. But DRP-Qos still can reach a relatively desirable accuracy (about 0.75)  since it pays less attention to nodes whose packets are more likely to lose efficacy.
        \begin{figure}[t]
            \centering
            \setlength{\belowcaptionskip}{-0.5cm}
            \includegraphics[height=5.2cm, width=0.35\textwidth]{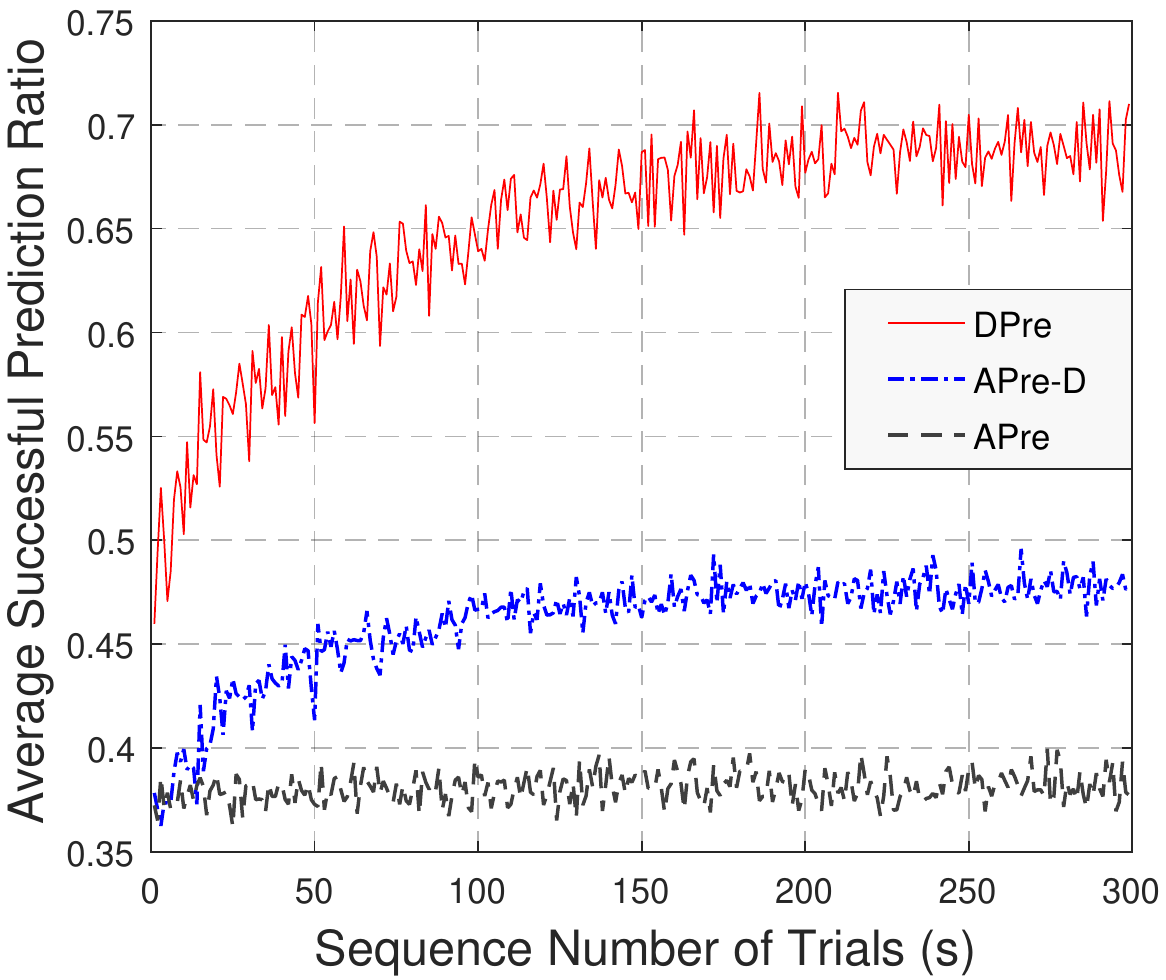}
            \vspace{-2pt}
            \caption{The prediction accuracy of DPre. }
            \label{total}
            \vspace{-4pt}
    \end{figure}
      \subsubsection{The DPre Framework Results}
     Finally, we evaluate our propose framework as a whole.  Fig~\ref{total} delineates the improvement brought by DPre (with differential utilities) under high dynamics and interference. As a baseline, APre represents the algorithm in~\cite{Predictive} that just reserves adjacent nodes for candidates. Note that APre does not consider the resources constraint, and thus we only uniformly distribute $Nres$ to each triggered candidate. To show the performance without the static stage, we combine APre with dynamic learning process, in which static reserve set collection is simplified to just choose the adjacent nodes (Here, $\xi = 8$ in both DPre and APre-D). It can be seen that without node filtering, APre achieves the lowest accuracy, which has no improvement. But through sequential learning, APre-D slightly improves its prediction accuracy since it would focus on a few correlated nodes in the static reservation sets after exploration. However, the high percentage of interference nodes impedes its accuracy improvement, that is also the reason why APre-D and APre perform much worse than DPre.
   \vspace{-2pt}
   \section{Related Work}~\label{relatedwork}
   5G has received increasing attention in industrial automation due to deterministic medium access.
   To provide critical QoS guarantees based on SPS, a series of massive access management (MAM) based on clustering are proposed in~\cite{AGTI}\cite{AGTILQ}\cite{AGTI3}. MTC devices in the same cluster are only allowed to access within an allocated access grant time interval (AGTI). In~\cite{AGTI}, the service rate of AGTIs is computed to ensure enormous QoS characteristics of different constant-rate MTC clusters. To guarantee statistical delay requests, a scheme is proposed in~\cite{AGTILQ} to get the minimum constant service rate through effective bandwidth. Due to the unpredictable nature of sporadic traffic in factory, an adaptive MAM in~\cite{AGTI3} utilizes historical observations to allocate AGTIs thus avoiding any reliance on prior stochastic assumptions.
   However, previous AGTI-based works overlook the fact that AGTIs are too scarce to allocate for each device. Also, we cannot ensure that devices in a cluster will be triggered concurrently within an AGTI.

   From another view, the works in~\cite{ASPS}\cite{delegation} aim to increase resource utilization considering the coarse granularity of resources allocated for individual node. The authors in~\cite{ASPS} propose an adaptive SPS scheme to adjust the resources allocated in the next transmission via buffer reports. To further utilize unused resources,
    resource delegation as a promising technique can increase overall throughput by leveraging from D2D~\cite{delegation}, where partially unused scheduling grants that were originally assigned in a semi-persistent manner could be granted to other devices.
    However, additional access delays may be brought in with extra D2D and buffer information exchange. Moreover, due to the small data transmissions feature in factory, BS can support an unique small resource size for MTC devices so that packets can be transmitted within a RB~\cite{AGTI}.

    To take both latency and resource utilization into account, a predictive uplink resource allocation scheme is introduced in~\cite{Predictive} for event monitoring applications, where correlated traffic characteristics are exploited to proactively assign uplink grants to devices in lieu of waiting for them to reactively request.
    However, without a precise prediction strategy, it will deteriorate the resource utilization due to the wrong pre-allocation decision. Hence, pre-allocation is the key to low-latency guarantees but it must be in place to ensure efficiency.
    \vspace{-13pt}
\section{Conclusion}\label{conclusion}
    \vspace{0pt}
    In this paper, we propose DPre, a predictive pre-allocation framework to explore the correlation between uplink traffics in industrial process automation, which is then exploited to make pre-allocation decisions to reduce access latency for delay-sensitive applications. Through supervised and sequential learning, DPre can significantly improve prediction accuracy and thus maximize the rewards of reserved resources.
 \vspace{-10pt}
\bibliographystyle{IEEEtran}
\nocite{*}\bibliography{IEEEabrv,infocom}

\end{document}